\documentclass[12pt,aps,reprint,groupedaddress,showpacs,nofootinbib,prd]{revtex4-1}
\usepackage{amsmath,amssymb,mathrsfs,graphicx,slashed,ragged2e,amsbsy}
\usepackage{enumitem}  
\usepackage{hyperref}

\newcommand{\bv}{\boldsymbol{v}}
\newcommand{\bb}{\boldsymbol{B}}
\newcommand{\EGW}{\dot{E}_{\text{GW}}}
\newcommand{\rS}{R_{\star}}

\begin{document}

\title{Monopolar and quadrupolar gravitational radiation from magnetically deformed neutron stars in modified gravity}

\author{Arthur George Suvorov}
\email{suvorovarthur@gmail.com}
\affiliation{School of Physics, University of Melbourne, Parkville VIC 3010, Australia}

\date{\today}

\begin{abstract}
Some modified theories of gravity are known to predict monopolar, in addition to the usual quadrupolar and beyond, gravitational radiation in the form of `breathing' modes. For the same reason that octupole and higher-multipole terms often contribute negligibly to the overall wave strain, monopole terms tend to dominate. We investigate both monopolar and quadrupolar continuous gravitational radiation from neutron stars deformed through internal magnetic stresses. We adopt the Parameterised-Post-Newtonian formalism to write down equations describing the leading-order stellar properties in a theory-independent way, and derive some exact solutions for stars with mixed poloidal-toroidal magnetic fields. We then turn to the specific case of scalar-tensor theories to demonstrate how observational upper limits on the gravitational-wave luminosity of certain neutron stars may be used to place constraints on modified gravity parameters, most notably the Eddington parameter $\gamma$. For conservative, purely poloidal models with characteristic field strength given by the spindown minimum, upper limits for the Vela pulsar yield $1-\gamma \lesssim 4.2 \times 10^{-3}$. For models containing a strong toroidal field housing $\sim 99\%$ of the internal magnetic energy, we obtain the bound $1- \gamma \lesssim 8.0 \times 10^{-7}$. This latter bound is an order of magnitude tighter than those obtained from current Solar system experiments, though applies to the strong-field regime. 

\end{abstract}


\maketitle

\section{Introduction}


Gravitational wave (GW) detectors, such as The Laser Interferometer Gravitational-Wave Observatory (LIGO) \cite{ligo}, are continually improving in sensitivity. It is therefore of ever-increasing importance to better understand what we can learn from further detections of GWs from astrophysical bodies. To this end, GW astronomy has two major aims: (i) to study the properties of bulk matter in astrophysical environments (including at high redshifts) \cite{nstest1,nstest2}, and (ii) to probe the theory of general relativity (GR) \cite{grtest1,grtest2}. For objects such as neutron stars, however, where extreme nuclear matter and strong-gravity coexist, there is a fundamental inseperability between these two aims \cite{comp1x,comp2,comp3,comp1xx}. For example, {a theoretical upper limit on the maximum mass of a neutron star is set by the Tolman-Oppenheimer-Volkoff relations} \cite{muxp,ma55}, the particulars of which intimately involve the equation of state of neutron matter \cite{tovx} and the gravitational action \cite{tov1,tov2}. {The observational existence of high mass neutron stars}, such as J$0348$+$0432$ $(M_{\star} = 2.01 \pm 0.04 M_{\odot})$ \cite{anton}, {therefore challenges} theories of gravity which do not permit such a mass for realistic equations of state \cite{maxmas1,maxmas2} or which suffer from instabilities for $M_{\star} \gtrsim 2 M_{\odot}$ \cite{unstable2}. More generally, to make full use of GW data, it is pertinent to identify what signatures, if any, a given theory of gravity imprints on a GW signal.

It is well known that GWs are generated by a rotating, biaxial object when the angle made between its angular momentum and symmetry axis vectors (`wobble angle') is non zero \cite{brady98,cut02}. In a theory of gravity wherein Birkhoff's theorem does not hold \cite{birk}, even stars pulsating \cite{weakd1,weakd2} or collapsing \cite{scheel,ott15} spherically can emit GWs through the so-called `breathing' mode \cite{breathe}. In a scalar-tensor theory, for example, energy can be dynamically exchanged between the scalar and tensor sectors through Yukawa-like interactions \cite{yukawa}, leading to the excitation of a breathing mode, whose contributions start at the post-Newtonian monopole level, and to a modulation of the usual $+$ and $\times$ tensor modes, whose contributions begin at the Newtonian quadrupole level \cite{scalarten,lee74,eard74,sotani}. For a compact object with spin frequency $\nu$ and radius $R_{\star}$, simple estimates show that the contribution to the overall GW strain from each $\ell$-pole scales with $\left(\nu R_{\star}/c\right)^{\ell}$ \cite{thorne80}. For a neutron star, this dimensionless ratio is of the order $\nu R_{\star} / c \lesssim 10^{-2}$ \cite{tovx}; a testament to the accuracy of the famous quadrupole formula. However, in a theory of gravity which permits it, this same scaling relationship implies that the monopole breathing mode {may} dominate over the quadrupole-generated mode(s). 

In any case, we focus here on magnetically deformed neutron stars as {potential} continuous GW sources. The equatorial magnetic field strength $B_{\star}$ of a neutron star can be estimated from its spindown, viz. $B_{\star} \approx 3.2 \times 10^{19} \sqrt{ P \dot{P}} \text{ G}$ for spin period $P$ (e.g. \cite{tovx}). Furthermore, measurements of cyclotron resonant scattering lines \cite{cyc1} and pulse fractions in surface X-ray emissions \cite{xray} suggest that some neutron stars have local magnetic field strengths well in excess ($\gtrsim$ 3 orders of magnitude in some cases, such as 1E $1207.4$-$5209$ \cite{psrb}) of their spindown limits. Since magnetic stresses are known to induce mass density asymmetries within a star \cite{chanf53,goos72}, magnetised neutron stars are therefore expected, {in principle}, to be excellent sources of continuous GWs \cite{bonaz96,laskrev}. In particular, these density asymmetries naturally lead to the generation of mass multipole moments, whose magnitude is proportional to the magnetic energy \cite{cut02,hask08,mmra11}. However, the precise relationship between the GW luminosity and the multipole moments depends on the theory of gravity \cite{thorne80,quadbed,landgen}. This makes a general assessment of modified-gravity-related consequences for GW emission from magnetically (or otherwise) deformed neutron stars a challenging task.

To make headway in a manner which is, at least somewhat, theory-independent, we introduce a novel approach based on the Parameterised-Post-Newtonian (PPN) formalism \cite{ppnX,ppnNI,ppn1,ppn2}. A natural generalisation of the post-Newtonian expansion \cite{pn1,blanchet,greenberg}, the PPN formalism provides a framework to quantify the impact of modified gravity parameters by building a generic metric which includes, to leading order, all possible geometric responses to material stresses \cite{ppnmhd1,ppnmhd2}. In theories of gravity which abide by the (Einstein) equivalence principle, wherein matter and otherwise non-gravitational fields couple only to the metric, ten such independent terms can emerge \cite{ppn1,ppnreview}. Each of these constituent pieces appear in a PPN `super-metric' with some coupling coefficients (the PPN parameters), which are to be constrained by experiment. These parameters take definitive values in a given theory of gravity \cite{ppn2} (see also Sec. II. B), though can be treated as free so that one may probe multiple theories simultaneously. For example, one of these coefficients is the classical Eddington one, denoted $\gamma$, originally used to parameterise the extent of light deflection by gravitational sources as a means to test GR \cite{eddington}, which predicts $\gamma =1$ exactly. Incidentally, the parameter $\gamma$ is also linked to the possibility of monopolar radiation \cite{sexl66}; see Sec. V.

Applying the PPN theory to the magnetohydrodynamic (MHD) study of neutron stars (Secs. II and III), their leading-order properties can be described in terms of the PPN parameters (Sec. IV); see also Refs. \cite{wagoner74,glampap15}. This allows us to investigate how modified gravity terms regulate the GW luminosity (Sec. V. A), which, upon comparison with LIGO and Virgo upper limits, effectively allows for constraints to be placed on modified gravity and stellar parameters (Sec. V. B). We restrict our attention to axisymmetric stars with mixed poloidal-toroidal, dipolar magnetic fields for simplicity, though we present the formalism in a general manner.

\section{Stellar structure}


For a neutron star, the magnetic energy density is $\lesssim 10^{-6} \left(B_{\star} / 10^{15} \text{ G}\right)^{2}$ that of the gravitational binding density. Therefore, even for anomalous X-ray pulsars or soft gamma repeaters (`magnetars') \cite{duncan}, we may treat the (post-Newtonian) Lorentz force as a perturbation on a background hydrostatic equilibrium \cite{mmra11,lasky13,mast16}. Handling magnetic forces as perturbations introduces non-barotropic features into the fluid, which are expected to emerge on physical grounds; see Refs. \cite{r92,nbarox}. As such, here we describe the PPN formalism, both in general (Sec. II. A) and for specific modified theories of gravity (Sec. II. B), and write down the relevant equations of motion for an unmagnetised star in equilibrium (Sec. II. C). Magnetic perturbations are then introduced in Sec. III.


\subsection{Parameterised-Post-Newtonian formalism}

In most modern theories of gravity, the metric tensor is a fundamental dynamical variable. For any given matter source, the field equations determine the metric structure. Because the theory space for modified gravity models is so vast, techniques to decompose a general metric are useful since they allow one to consider multiple classes of theories at once. In this paper, we adopt the PPN formalism \cite{ppn1,ppn2}, a review of which can be found in Ref. \cite{ppnreview}. Different approaches, such as Geroch-Hansen moment deconstructions \cite{multGER,mult1,mult2,mult3} ({which expand a metric as a sum over moments rather than post-Newtonian terms}), require spacetime symmetries which will often not be present when generic matter couplings are considered (e.g. non-stationary sources).

For our purposes, it is sufficient to note that the PPN formalism allows one to expand the components of a general\footnote{Strictly speaking, the formalism applies to theories wherein (i) the metric tensor is symmetric, (ii) test bodies follow Levi-Civita geodesics, and (iii) special relativity describes non-gravitational physics in freely falling frames \cite{ppnmhd1,ppn2}.} metric tensor into powers of dimensionless quantities such as $v^2/c^2$, where $c$ is the speed of light and $v \ll c$ is the velocity of the source, thus encapsulating the dynamics of slow-motion sources up to some desired order. This is similiar to the post-Newtonian expansion of GR \cite{blanchet}, though applies to more general metric theories of gravity. Once one has identified all possible terms which can appear in the metric at post-Newtonian order, a `super metric'  is built which includes all such terms coupled to some coefficients which are \textit{a priori} arbitrary \cite{ppnmhd1}. These coefficients are, however, uniquely determined in any given theory of gravity; see chapter 5 of Ref. \cite{ppn2}.

In general, the PPN metric contains 10 coefficients\footnote{Throughout this work, we adopt the notation first introduced by Will in Ref. \cite{ppnmhd1}, which is slightly different to the standard `alpha-zeta' notation, but is more compact; compare equations (16) of \cite{ppnmhd1} with those on page 31 of \cite{ppn2} to translate between the two notations. Note that we have restored dimensional constants in our expressions.}: $\gamma$ and $\beta$, which are the classical Eddington-Robertson-Schiff parameters \cite{eddington} quantifying spatial curvature due to unit mass (e.g. light deflection), $\zeta$ and $\Sigma$, quantifying spatial curvature due to the radial and transverse components of kinetic energy and stress (if different from unity, indicates preferred-frame or Whitehead effects \cite{whitehead}), $\Delta_{1}$ and $\Delta_{2}$, describing how much inertial frames are dragged by unit momentum (if different from unity, indicates a violation of conservation of angular momentum), and finally $\beta_{i}$ for $i=1,2,3,4$, quantifying spatial curvature due to unit kinetic energy, unit gravitational potential, unit internal energy, and unit pressure, respectively (if different from unity, indicates a violation of conservation of linear momentum). One can show that, under an appropriate gauge fixing, the coefficient $\Sigma$ is mathematically redundant \cite{ppn1}. Without loss of generality, we henceforth set $\Sigma = 0$. In GR, all parameters have unit value except for $\zeta$, which vanishes. 

In general, the PPN metric reads
 \begin{widetext}
\begin{equation} \label{eq:ppnmetric}
ds^2 =  c^2\left( 1 - \frac {2 U}{c^2} + \beta \frac {2U^2}{c^4} - \frac{4\Phi}{c^2} + \zeta \mathcal{A} \right) d t^2 + \left( \frac {7} {2} \Delta_{1} V_{a} + \frac {1} {2} \Delta_{2} W_{a} \right) dt dx^{a} - \left( 1 + \frac{2 \gamma U} {c^2} \right) \delta_{ab} dx^{a} dx^{b},
\end{equation}
\end{widetext}
where $U$ is the standard Newtonian gravitational potential satisfying 
\begin{equation} \label{eq:pois1}
\nabla^2 U = -4 \pi G \rho,
\end{equation}
$\nabla$ represents the usual flat space derivative operator, $G$ is Newton's constant, $\rho$ is the mass density, $\Phi$ is the generalised `post-Newtonian potential',
\begin{equation} \label{eq:pois2}
\nabla^2 \Phi = -4 \pi G \rho \left( \beta_{1}\frac {v^2}{c^2} + \beta_{2} \frac{U}{c^2} + \beta_{3}\frac {\Pi} {2} + \beta_{4}\frac {3 p} {2c^2 \rho} \right),
\end{equation}
for pressure $p$ and internal energy density $\rho \Pi$, $\mathcal{A}$ is given through the Green's function expression
\begin{equation}
\mathcal{A}(\boldsymbol{x},t) = \frac {G} {c^4} \int d \boldsymbol{x}' \frac {\rho(\boldsymbol{x}',t) \left[\boldsymbol{v}(\boldsymbol{x}',t) \cdot \left( \boldsymbol{x} - \boldsymbol{x'} \right) \right]^{2}} {|\boldsymbol{x} - \boldsymbol{x}'|^{3}} ,
\end{equation}
and the vector potentials $\boldsymbol{V}$ and $\boldsymbol{W}$ are similarly defined via
\begin{equation}
\boldsymbol{V}(\boldsymbol{x},t) =  \frac{G}{c^2} \int d \boldsymbol{x}' \frac {\rho(\boldsymbol{x}',t) \bv(\boldsymbol{x}',t)} {|\boldsymbol{x} - \boldsymbol{x}'|} ,
\end{equation}
and
\begin{equation} \label{eq:wpotential}
\boldsymbol{W}(\boldsymbol{x},t) = \frac{G}{c^2} \int d \boldsymbol{x}' \frac {\rho(\boldsymbol{x}',t) \left( \boldsymbol{x} - \boldsymbol{x'} \right) \left[ \bv(\boldsymbol{x}',t) \cdot \left( \boldsymbol{x} - \boldsymbol{x'}\right) \right] } {|\boldsymbol{x} - \boldsymbol{x}'|^{3}} ,
\end{equation}
respectively. 

\subsection{PPN coefficients in modified gravity}

In a general theory of gravity, the PPN coefficients are not constant but vary as functions of the Ricci scalar $R$ or other curvature quantities \cite{ppnpars1,ppnpars2,ppnpars3}. For example, in massive Brans-Dicke theories \cite{ppnparsX}, the PPN parameters scale inversely with the distance from compact sources because the scalar field becomes suppressed in regions of weak curvature through the Vainshtein \cite{vain1} or other mechanisms \cite{cham1}. This is important because it implies that Solar system constraints (see below) on the PPN parameters are not necessarily applicable to the neutron star regime. Indeed, there are only a few strong field constraints on the PPN parameters, which come from pulsar timing \cite{will92,bell,stairs}. Nevertheless, accurate measurements of $P$ and its derivatives from a pulsar system strongly constrains deviations from momentum conservation (mainly $\beta_{1}$ and $ \beta_{2}$) and the existence of preferred frame effects $(\zeta)$ at high energies, and somewhat constrains other parameters (though see Ref. \cite{esposito}). In theories where the PPN parameters are everywhere constant (e.g. massless Brans-Dicke theory), non-Einstein terms negligibly affect fluid behaviour because the PPN parameters are necessarily heavily constrained from both the weak and strong field experiments \cite{ppnreview}. 

To provide an explicit example of the above considerations, it can be shown that for the $f(R)$ theory of gravity (see Ref. \cite{fofrreview} for a review), the PPN parameters $\gamma$ and $\beta$ take the form \cite{ppnfr,ppnfr2}
\begin{equation} \label{eq:frgamma}
\gamma -1 = - \frac {f''(R)^2} {f'(R) + 2 f''(R)^2},
\end{equation}
and
\begin{equation} \label{eq:frbeta}
\beta - 1 = 	\frac {f'(R) f''(R)} {8 f'(R) + 12 f''(R)^2} \frac {d \gamma} {d R} ,
\end{equation}
respectively. Note that, in the GR limit, $f(R) = R$, the above formulas yield $\gamma = \beta = 1$ exactly, as expected. The Cassini spacecraft \cite{cassini} and lunar laser ranging (Nordtvedt effect) \cite{lunar} experiments imply the constraints \cite{ppnreview}
\begin{equation} \label{eq:solar1}
|\gamma_{0} - 1 | \leq 2.3 \times 10^{-5}, \,\,\,\,\,\,\, |\beta_{0} - 1| \leq 2.3 \times 10^{-4},
\end{equation}
where the subscripts indicate that the PPN parameters are evaluated at the Solar system value of the Ricci scalar, $R_{0} \approx 10^{-27} \text{ cm}^{-2}$ \cite{comp1x}. Simple estimates, based on the fact that $R \propto G \rho / c^{2}$ \cite{ric2}, show that the scalar curvature in a neutron star core exceeds the Solar system value by many orders of magnitude; $R_{\text{NS}} \approx 10^{-12} \text{ cm}^{-2}$. 

\begin{figure}[h!]
\includegraphics[width=0.473\textwidth]{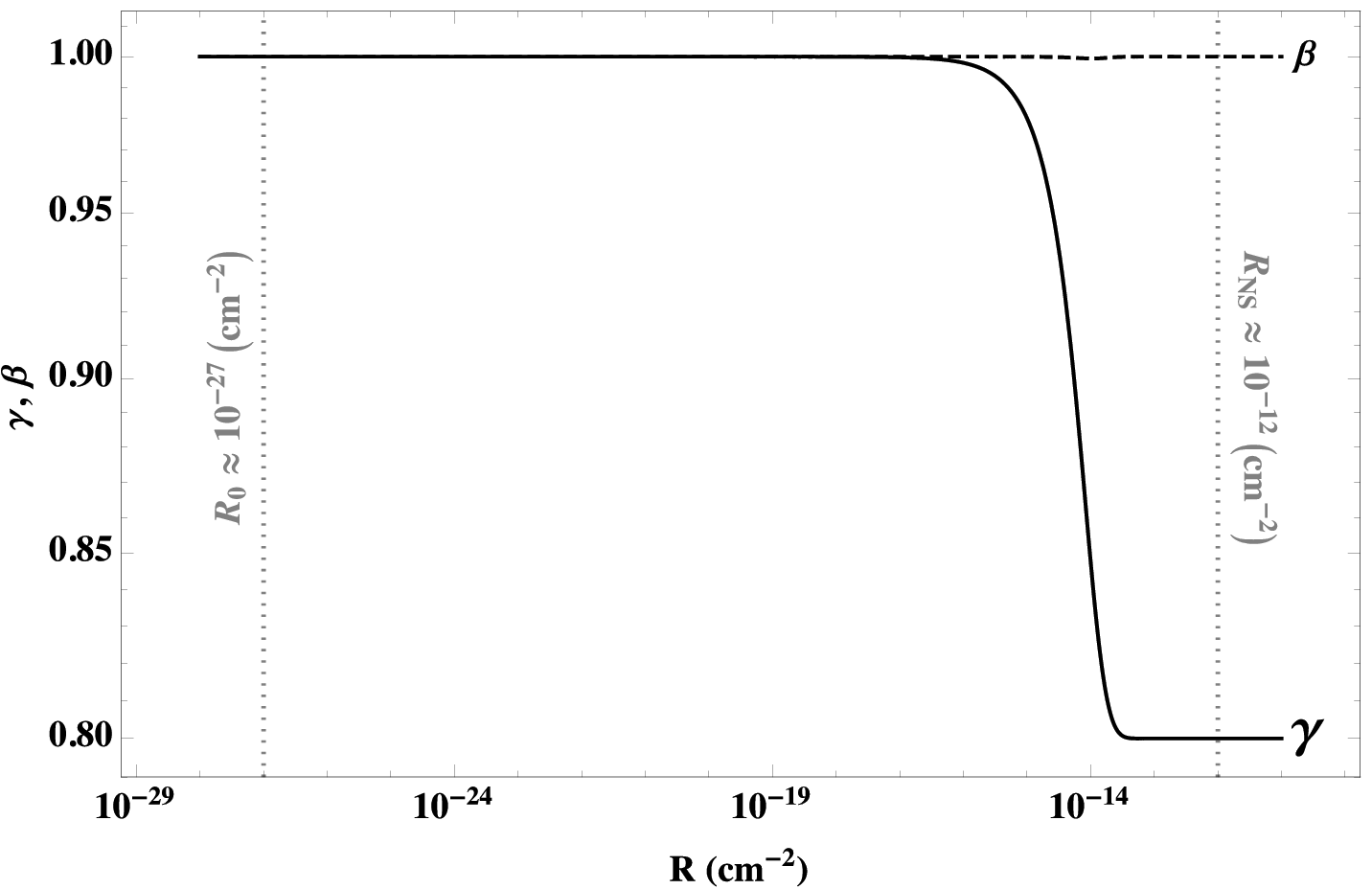}
\caption{Values of the PPN parameters $\gamma$ (solid curve) and $\beta$ (dashed curve) as functions of the Ricci scalar $R$, evaluated through \eqref{eq:frgamma} and \eqref{eq:frbeta}, respectively, for a particular $f(R)$ theory \eqref{eq:particularf}. Approximate values of the scalar curvature in the Solar system $R_{0} \approx 10^{-27} \text{ cm}^{-2}$ and inside a neutron star core $R_{\text{NS}} \approx 10^{-12} \text{ cm}^{-2}$ are shown by dotted vertical lines. \label{gammabeta}
}
\end{figure}

In Figure \ref{gammabeta} we show the PPN parameters $\gamma$ [equation \eqref{eq:frgamma}, solid curve] and $\beta$ [equation \eqref{eq:frbeta}, dashed curve] for an $f(R)$ theory of gravity given by
\begin{equation} \label{eq:particularf}
f(R) = \lambda - \frac {1} {4} \int d R \left[ \int d R \left( \frac { Q(R) - 1} {2Q(R) - 1} \right)^{1/2} \right]^{2},
\end{equation}
where $\lambda$ is an arbitrary (cosmological) constant and 
\begin{equation}
Q(R) = 1 - \frac {1} {5} \tanh \left( \frac {R} {10^{13} R_{0}} \right).
\end{equation} 
Note that this choice for the function $f$ \eqref{eq:particularf} is not motivated by observational considerations (though see Ref. \cite{tanhref}), but is chosen to demonstrate that the PPN coefficients can take values different from unity in a neutron star regime, even when Solar system experiments \eqref{eq:solar1} are accounted for and a viable weak-field limit exists. Indeed, we see that the constraints \eqref{eq:solar1} are satisfied at $R \sim R_{0}$, while the parameter $\gamma$ departs from one when $R \gg R_{0}$, eventually saturating at $\gamma = 0.8$ for extreme scalar curvatures $R \gtrsim 10^{-2} R_{\text{NS}} \gg R_{0}$. The parameter $\beta$ is everywhere less than one, though departs negligibly from unity for $R \sim R_{0}$ and by $\lesssim 1$ percent across all values of $R$. Similar behaviours for the PPN parameters can occur in other theories of gravity, e.g. massive Brans-Dicke theories \cite{ppnparsX}.

\subsection{PPN hydrodynamics}

Using expressions \eqref{eq:ppnmetric}--\eqref{eq:wpotential}, one can derive\footnote{In presenting these equations and throughout the rest of the paper, we implicitly assume that the PPN coefficients vary slowly over the length and energy scales associated with the star, i.e. we assume $\nabla \gamma \ll \mathcal{O}(c^{-2})$ and similarly for all other PPN parameters.} the PPN-continuity equation \cite{ppnmhd1},
\begin{equation} \label{eq:continuity}
0 = \frac {\partial \rho^{\star}} {\partial t} + \nabla \cdot \left( \rho^{\star} \boldsymbol{v} \right),
\end{equation}
and the PPN-Euler equations,
\begin{widetext}
\begin{equation} \label{eq:euler}
\begin{aligned}
0 =& \rho^{\star} \frac {d \bv} {d t} - \rho^{\star} \nabla U + \nabla \left[ p \left( 1 + \frac{3 \gamma U} {c^2} \right) \right] - \left( \frac {v^2} {2 c^2} + \Pi + \frac{p}{c^2\rho^{\star}} \right) \nabla p + \rho^{\star} \frac {d} {dt} \left[ \frac{\left(2 \gamma + 2 \right)U}{c^2} \bv  - \frac {1} {2} \left( 7 \Delta_{1} + \Delta_{2} \right) \boldsymbol{V} \right] \\
& - \frac {\bv} {c^2} \left( \rho^{\star} \frac {\partial U} {\partial t} - \frac {\partial p} {\partial t} \right) - \frac {1} {2} \Delta_{2} \rho^{\star} \frac {\partial} {\partial t} \left( \boldsymbol{W} - \boldsymbol{V} \right) + \frac {1} {2} \left( 7 \Delta_{1} + \Delta_{2} \right) \rho^{\star} \bv \cdot  \nabla \boldsymbol{V} + \frac {1} {2} \zeta \rho^{\star} c^2 \nabla \mathcal{A} \\
& - 2 \rho^{\star} \nabla\Phi - \rho^{\star} \left[ \gamma v^2 - \left(2 \beta - 2 \right) U + \frac{3 \gamma p} {\rho^{\star}} \right] \frac{\nabla U} {c^2} ,
\end{aligned}
\end{equation}
\end{widetext}
where $d/dt = \partial / \partial t + \bv \cdot \nabla$ is the advective derivative, and we have introduced the so-called `conserved density'
\begin{equation}
\rho^{\star} = \rho \left( 1 + \frac {1} {2} \frac{v^2}{c^2} + \frac{3 \gamma U}{c^2} \right).
\end{equation}

Equations \eqref{eq:continuity} and \eqref{eq:euler}, subject to an equation of state relating $p$ to $\rho$ and other thermodynamic variables, completely characterise the properties of a perfect fluid at first post-Newtonian order in a metric theory of gravity (see also Ref. \cite{bianchi}). Upon discarding terms of order $\mathcal{O}(c^{-2})$, one recovers Newtonian hydrodynamics.

In order to simplify the calculations and subsequent analysis, we use a static approximation for the star wherein $\bv = 0$ and $\partial_{t} = 0$. In this case, the continuity equation \eqref{eq:continuity} is identically satisfied, and the Euler equation \eqref{eq:euler} is greatly simplified. We also ignore thermodynamic contributions to the total density (e.g. compressional or thermal energies); $\Pi=0$. Ultimately, this means that we need only consider four of the PPN parameters, namely $\gamma, \beta, \beta_{2}$, and $\beta_{4}$, as the others do not enter into the equations of motion. Including thermodynamic and kinematic effects is, in principle, straightforward, but requires a more sophisticated numerical analysis than is employed here (see Sec. IV. A).

\section{Magnetised stellar structure}

The Lorentz force $\boldsymbol{F}^{L}$, to $\mathcal{O}(c^{-2})$ in a metric theory of gravity, is given by \cite{ppnmhd2}
\begin{equation} \label{eq:lorentz}
\boldsymbol{F}^{L} = \frac {1} {\mu_{0}} \left[ \nabla \times \left( \varphi \bb \right) \right] \times \bb,
\end{equation}
where $\varphi = 1 - 2\gamma U/c^2$, $\bb$ is the magnetic field, and $\mu_{0}$ represents the permeability of free space. Throughout the rest of this work we adopt spherical coordinates $(r,\theta,\phi)$.

We model a magnetised neutron star by introducing an axisymmetric perturbation $\rho \rightarrow \rho + \delta \rho(r,\theta)$ and $p \rightarrow p + \delta p(r,\theta)$ into \eqref{eq:euler} such that the force \eqref{eq:lorentz} acts as the source of the perturbation, as in Refs. \cite{mmra11,lasky13,mast16}. We assume no \textit{a priori} relationship between $\delta \rho$ and $\delta p$ (hence non-barotropic). We further ignore self-gravity corrections, i.e. we impose the Cowling and `post-Cowling' conditions that $\delta U$ and $\delta \Phi$ vanish identically while ignoring the perturbed versions of equations \eqref{eq:pois1} and \eqref{eq:pois2}. We obtain a simple set of perturbation equations from \eqref{eq:euler}
\begin{equation} \label{eq:eulerpert}
\begin{aligned}
0=&  \left(1 + \frac {3 \gamma U} {c^2} \right)\ \left[ \nabla \delta p - \delta \rho \nabla \left( U + 2 \Phi + \frac {1 - \beta} {c^2} U^2 \right) \right]  \\
&+ \frac {\delta p \nabla U} {c^2} \left( 1 + \frac{3 \gamma U}{c^2} \right)^{-1} - \boldsymbol{F}^{L} ,
\end{aligned}
\end{equation}
which reduce to the standard MHD perturbation equations when terms of order $\mathcal{O}(c^{-2})$ are discarded \cite{mmra11,lasky13,mast16}. The perturbed continuity equation \eqref{eq:continuity} is identically satisfied. 

Equation \eqref{eq:eulerpert} describes the structure of a nonbarotropic, magnetised star to leading order in the modified gravity parameters $\gamma, \beta, \beta_{2}$, and $\beta_{4}$. Note that the magnetic field is subject to the usual divergence free condition $\nabla \cdot \bb = 0$ even at post-Newtonian level \cite{greenberg}.

\subsection{Axisymmetric magnetic fields}

In this paper, we consider axisymmetric magnetic fields for simplicity. In this case, one can decompose the vector $\bb$ into a sum of poloidal and toroidal components \cite{chand56},
\begin{equation} \label{eq:bfield}
\bb = B_{0} \left[ \nabla \psi \times \nabla \phi + \left(\frac {E_{p}} {E_{t}} \frac {1 - \Lambda} {\Lambda} \right)^{1/2} F \nabla \phi \right],
\end{equation}
where $B_{0}$ is the characteristic field strength, $\psi = \psi(r,\theta)$ is the scalar flux function, and $F = F(r,\theta)$ describes the spatial variation of the azimuthal (toroidal) component of the field. In expression \eqref{eq:bfield}, we have introduced the quantities 
\begin{equation} \label{eq:epol}
E_{p} = \frac {B_{0}^2} {2 \mu_{0}} \int_{V} dV \left[ \left( \frac {1} {r^2 \sin\theta} \frac {\partial \psi} {\partial \theta} \right)^2 + \left( \frac {1} {r \sin\theta} \frac {\partial \psi} {\partial r} \right)^2 \right],
\end{equation}
and
\begin{equation} \label{eq:etor}
E_{t} = \frac {B_{0}^2} {2 \mu_{0}} \int_{V} dV  \frac {F^2} {r^2 \sin^2\theta},
\end{equation}
which represent the energies, stored within stellar volume $V$, of the poloidal and toroidal components of the magnetic field, respectively. In particular, the toroidal prefactor $\left(\frac {E_{p}} {E_{t}} \frac {1 - \Lambda} {\Lambda} \right)^{1/2}$ in \eqref{eq:bfield}, where $0 < \Lambda  \leq 1$, characterises the relative strength between the poloidal and toroidal components, i.e. $\Lambda = 0.1$ gives a field for which $90\%$ of the magnetic energy is stored within the toroidal field, $\Lambda = 0.5$ gives a field which has an equal poloidal-to-toroidal field strength ratio $E_{p} = E_{t}$, and $\Lambda =1$ gives a purely poloidal field.

By virtue of axisymmetry, the azimuthal component of the PPN-Lorentz force \eqref{eq:lorentz} must vanish. This can be achieved only if the function $F$ defined in \eqref{eq:bfield} behaves in a certain way. In ordinary MHD, this amounts to requiring $F = F(\psi)$ \cite{chand56}. In the PPN case, the $\phi$-component of the Lorentz force \eqref{eq:lorentz} reads 
\begin{equation} \label{eq:lorazi}
0 = F \left( \frac {\partial \varphi} {\partial r} \frac{ \partial \psi} {\partial \theta} -  \frac {\partial \varphi} {\partial \theta} \frac {\partial \psi} {\partial r} \right) + \varphi \left( \frac {\partial F} {\partial r} \frac {\partial \psi} {\partial \theta} - \frac {\partial F} {\partial \theta} \frac {\partial \psi} {\partial r} \right).
\end{equation} 
Through a minor abuse of notation, we may solve \eqref{eq:lorazi} in general to obtain
\begin{equation} \label{eq:torreln}
F = \frac {F(\psi)} {\varphi(r,\theta)},
\end{equation}
which, upon discarding terms of order $\mathcal{O}(c^{-2})$, reduces to the expected solution $F = F(\psi)$. In accord with previous works \cite{mmra11,lasky13,mast16}, we make the choice
\begin{equation} \label{eq:exptor}
F(\psi) =
  \begin{cases}
                                    \left( \psi - \psi_{c} \right)^2/ \rS^3 & \text{for $\psi \geq \psi_{c}$} \\
                                   0 & \text{otherwise,} 
  \end{cases}
\end{equation}
where $\psi_{c}$ is the value of $\psi$ that defines the last poloidal field line that closes inside the star. The simple form \eqref{eq:exptor} for $F$ ensures that the toroidal field is confined to the equatorial torus which is bounded by the last closed poloidal field line.

To ensure a physically reasonable magnetic field, the streamfunction $\psi$ defining the magnetic field \eqref{eq:bfield} is, in general, subject to the following conditions \cite{mmra11,akgun}.
\begin{enumerate}[label=(\roman*)]
\item {$\bb$ is continuous with respect to a (force-free) dipole field outside the star,}
\item {there are no surface currents; $$\boldsymbol{J} \equiv \frac {1} {\mu_{0}} \left[ \nabla \times \left( \varphi \bb \right) \right] = 0$$ at the stellar surface $\partial V$, and}
\item {the current density $\boldsymbol{J}$ is finite at the origin.}
\end{enumerate}

\section{Explicit solutions}
In this section we present explicit solutions to the perturbation equations \eqref{eq:eulerpert}.

\subsection{Background profiles}
Before solving the perturbation equations, we need to select appropriate background profiles. Throughout this work, we use the simple parabolic density profile
\begin{equation} \label{eq:parabolic}
\rho(r) = \rho_{c} \left[ 1 - \left(\frac {r} {R_{\star}} \right)^{2} \right],
\end{equation}
where $\rho_{c} = 15 M_{\star}/\left( 8 \pi R_{\star}^3\right)$ is the central density for stellar mass $M_{\star}$. The density \eqref{eq:parabolic} decreases monotonically with radius, finally decaying to zero at the stellar surface $r = R_{\star}$, and has vanishing derivative at the origin. The choice \eqref{eq:parabolic}, although simple, is therefore reasonably realistic, and has been used in several previous works on magnetic deformations of neutron stars \cite{mmra11,akgun,yoshida}. {Furthermore, it was shown in Ref. \cite{mmra11} that GW-related observables associated with magnetically deformed stars are largely independent, at least to order $\mathcal{O}(c^{0})$, from the exact form of $\rho$; see Figure 4 in Ref. \cite{mmra11}.}

For \eqref{eq:parabolic}, the potential $U$ can be solved for exactly from the Poisson equation \eqref{eq:pois1} to give
\begin{equation} \label{eq:potentialu}
U(r) = \frac {G M_{\star}} {8 \rS^5} r^2 \left( 3r^2- 10 \rS^2 \right).
\end{equation}
In general, the background pressure $p$ and potential $\Phi$ need to be solved for numerically, primarily because of the $\beta_{4}$ term which couples $p$ to $\Phi$ through \eqref{eq:pois2} and adds an additional degree of nonlinearity to the Euler equation \eqref{eq:euler}. Note that the background hydrostatic star is spherically symmetric, so $p$ and $\Phi$ are necessarily functions of the radial coordinate only. The boundary conditions to be applied are the standard ones, namely that $p(R_{\star})=0$ and $\Phi(0) = \Phi'(0)=0$ \cite{tovx,ppnmhd1}. Because of the polynomial nature of $\rho$ \eqref{eq:parabolic}, we can readily solve the coupled ODEs \eqref{eq:pois2} and \eqref{eq:euler} subject to the aforementioned boundary conditions using a straightforward orthogonal polynomial spectral solver in {MATHEMATICA} \cite{wolfram}.



\subsection{PPN-MHD solutions for dipolar magnetic field}

As it turns out, regardless of the particulars of the Lorentz force, the radial and angular components of the perturbation equations \eqref{eq:eulerpert} can be solved exactly to yield
\begin{equation} \label{eq:ppnpres}
\delta p = \frac {c_{0}(r) + r \int d \theta \boldsymbol{F}^{L}_{\theta}} {1 + 3 \gamma U/c^2},
\end{equation}
and
\begin{equation} \label{eq:ppnrho}
\delta \rho = c^2 \frac {  -c^2 \boldsymbol{F}^{L}_{r} + \frac {c^2 \delta p  \frac {d U} {d r}} {\left( c^2 + 3  \gamma U \right)} + \left( c^2 +3 \gamma U \right) \frac{\partial \delta p}{ \partial r}}  {\left( c^2 + 3 \gamma U \right) \left\{ \left[ c^2 + 2 \left( \beta - 1 \right) U \right] \frac{d U}{d r} + 2 c^2 \frac{d \Phi}{d r} \right\}},
\end{equation}
where $c_{0}(r)$ is an arbitrary function arising through integration, chosen so as to ensure that $\delta p$ is continuous across the toroidal boundary $\psi = \psi_{c}$.

For simplicity, in this paper we consider a dipolar magnetic field \cite{mmra11}, although an analysis involving more general magnetic fields could be readily obtained using the results of Refs. \cite{lasky13,mast16}. For a dipole field, the poloidal streamfunction $\psi$ takes the simple form
\begin{equation} \label{eq:dipole}
\psi(r,\theta) = \kappa(r) \sin^2\theta,
\end{equation}
for some function $\kappa$ subjected to conditions (i)--(iii) presented in Sec. III. A. In particular, since the toroidal field \eqref{eq:exptor} vanishes near the boundary and origin of the star, these conditions are satisfied for a polynomial $\kappa$ containing no constant or linear terms, provided that
\begin{equation} \label{eq:currentdensity}
0 = \kappa''(R_{\star}) - \frac{2 \gamma U'(R_{\star})} {c^2} \kappa'(R_{\star}) - \frac {2 \kappa(R_{\star})}{R_{\star}^2} = 0,
\end{equation}
$\kappa(R_{\star}) = R_{\star}^2$, and $\kappa'(R_{\star}) = -R_{\star}$. Using \eqref{eq:potentialu}, we find that a suitable choice for $\kappa$ satisfying the above requirements is
\begin{equation} \label{eq:kappa}
\begin{aligned}
\kappa(r) =& \frac {r^2} {8 \rS^4} \Big[ 15 r^4 - 42 r^2 \rS^2 + 35 \rS^4 \\
&+ \gamma \frac {2 G M_{\star} } {c^2 R_{\star}} \left(\rS^2 - r^2 \right)^{2} \Big] .
\end{aligned}
\end{equation}
One will note that there are infinitely many solutions for $\kappa$ (cf. Appendix A of Ref. \cite{lasky13}), and we have merely chosen one such possibility. However, our choice is unique in the sense that it is only one that recovers the Newtonian results of Ref. \cite{mmra11} when terms of order $\mathcal{O}(c^{-2})$ are discarded. For \eqref{eq:dipole} with \eqref{eq:kappa}, we have $\psi_{c} = \rS^2$. In our models, the equatorial field strength is equal to $B_{0}$, i.e. $B_{\star} = B_{0}$.

\begin{figure*}
\includegraphics[width=\textwidth]{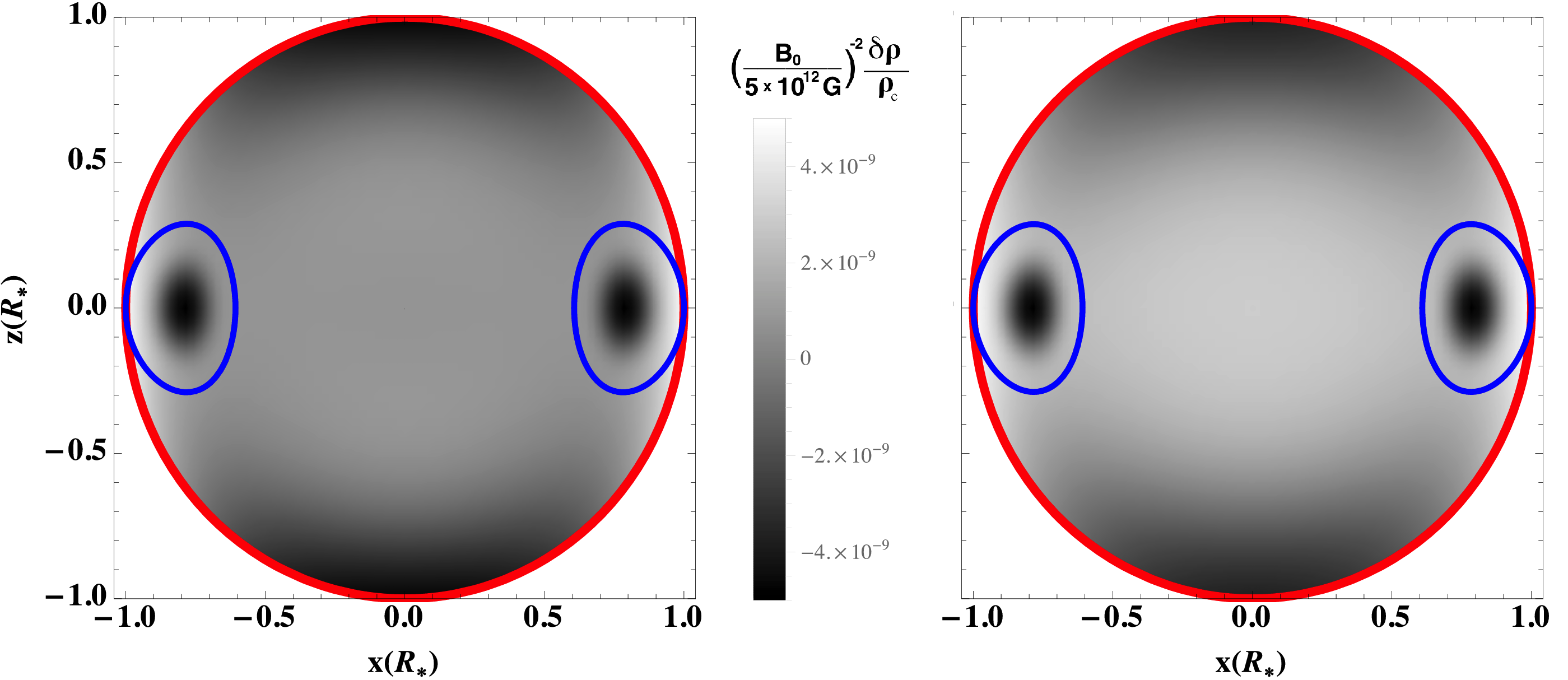}
\caption{Contours of the normalised density perturbation $\delta \rho / \rho_{c}$, given through expression \eqref{eq:ppnrho}, for $\gamma = \beta = \beta_{2} = \beta_{4} = 1$ (GR values; left panel) and $\gamma =0.6$, $\beta = 0.7$, $\beta_{2} = 1$, and $\beta_{4} = 1.2$ (right panel), poloidal-to-toroidal field strength ratio $\Lambda = 0.7$, stellar radius $R_{\star} = 10^{6} \text{ cm}$ and mass $M_{\star} = 1.4 M_{\odot}$. The stellar surface $r=\rS$ is shown in red, while the toroidal boundaries $\psi = \psi_{c}$ are shown in blue. \label{rho10}
}
\end{figure*}

We now have all the ingredients required to evaluate the solutions \eqref{eq:ppnpres} and \eqref{eq:ppnrho} as functions of the magnetic field parameters $B_{0}$ and $\Lambda$, the stellar parameters $M_{\star}$ and $R_{\star}$, and the PPN parameters $\gamma, \beta, \beta_{2}$, and $\beta_{4}$.
In Figure \ref{rho10} we show contours of the density perturbation \eqref{eq:ppnrho} subject to the Lorentz force \eqref{eq:lorentz} for the dipolar magnetic field \eqref{eq:dipole} with \eqref{eq:kappa}, where $\gamma=\beta=\beta_{2}=\beta_{4}=1$ (GR values; left panel) and $\gamma=0.6$, $\beta=0.7$, $\beta_{2}=1$, and $\beta_{4}=1.2$ (right panel), where $R_{\star} = 10^{6} \text{ cm}$, $M_{\star} = 1.4 M_{\odot}$, and $\Lambda =0.7$. The perturbed density profiles are qualitatively similar, displaying relatively large deformations of the order $\gtrsim 10^{-9} \left( B_{0}/5 \times 10^{12} \text{ G}\right)^{2} \rho_{c}$ in the toroidal $(\psi \geq \psi_{c})$ and polar cap $(r \gtrsim 0.9 R_{\star}, \theta \sim 0,\pi)$ regions. Furthermore, $\delta \rho$ is negative near the polar caps $(\theta \sim 0,\pi)$ and positive at the edges of the equatorial belt $(\theta \sim \pi/2)$ in both cases, indicating that the stars are oblate. This is expected because the poloidal field is globally stronger than the toroidal field for this simulation, i.e. $\Lambda > 0.5$ \cite{mmra11,lasky13,mast16} (see also Sec. V). The non-GR deformation is somewhat stronger $(\approx 35\%)$ than its GR counterpart in the toroidal regions because the toroidal function $F$ \eqref{eq:torreln} scales with $1/\varphi \approx 1 + 2 \gamma U /c^2$, which is larger in the non-GR case because $U \leq 0$. However, the non-GR deformation is weaker by a factor $\sim 2$ near the polar caps because the poloidal components of the Lorentz force \eqref{eq:lorentz} have the opposite scaling; $\boldsymbol{F}^{L}_{r,\theta} \propto \varphi$. In the core region $(r \lesssim 0.4 R_{\star})$, the non-GR deformation is slightly stronger $(\approx 10\%)$ because the denominator in expression \eqref{eq:ppnrho} is smaller for $\beta < 1$. The toroidal geometry varies slightly with different values of $\gamma$ because the function $\kappa$ \eqref{eq:kappa} depends on this parameter. However, for the range of stellar masses and radii considered here, the effect is small; the toroidal volume changes by $\lesssim 3\%$ between solutions with $0 \leq \gamma \leq 1.5$ for $R_{\star} \sim 10^{6} \text{ cm}$ and $M_{\star} \sim 1.4 M_{\odot}$.



\section{Gravitational waves}


Here we estimate the GW luminosity as a function of the PPN and stellar parameters for the stars described in Sec. IV.

\subsection{Multipolar radiation}

%

Loosely speaking, the gravitational radiation emitted by a source depends on the particulars of the `gravitational energy-momentum'. This latter quantity can be defined using the Landau-Lifshitz pseudotensor \cite{land1} and its non-Einstein generalisations \cite{nutku69,landgen,land2,nafjet}, or from quasilocal Hamiltonian constructions \cite{quasi2,quasi3}. In the former picture, the pseudotensor, when summed together with the material energy-momentum $\boldsymbol{T}$, forms a conserved current $\boldsymbol{\tau}$, from which one obtains a suitable definition of radiation energy. If there are non spin-2 fields within the theory which couple non-minimally, they may also contribute to the total energy-momentum (see below) \cite{nutku69,lee74,eard74}. In general, the structure of the object $\boldsymbol{\tau}$ depends on the specific form of the gravitational action \cite{landgen,ppn2,land1,thorne80}. Therefore, the details of both the radiation energy spectrum (namely $\boldsymbol{\tau}$) and the structure of magnetised stars [namely \eqref{eq:ppnpres} and \eqref{eq:ppnrho}] separately depend on the PPN coefficients in nontrivial ways. 


To make progress, we can consider a relatively simple but representative example theory in order to get an idea for how the radiation energy behaves for the stars described in Sec. IV. To this end, we consider a scalar-tensor theory, which, as previously mentioned, permits a breathing mode \cite{lee74,breathe}. That is, scalar-tensor theories allow for monopole and higher-order terms to be non-zero when one performs a multipole expansion for the radiation field. This can be thought of as a consequence of dissipative Yukawa-like interactions that are allowed to occur between the scalar and tensor sectors \cite{yukawa,weakd1,weakd2}. Omitting the details (which can be found in chapter 10.3 of Ref. \cite{ppn2} or section 2.3 of Ref. \cite{orient2}), one can define the conserved currents discussed above for scalar-tensor theories \cite{nutku69,eard74} and, upon keeping leading-order\footnote{In general, terms proportional to $(\beta-1)$ may appear within the multipole expansion of the radiation field, though are sub-leading in our case because we have assumed throughout (see Footnote 3) that $\nabla \gamma \ll \mathcal{O}(c^{-2})$; cf. expression \eqref{eq:frbeta}.} terms (i.e. Newtonian terms up to quadrupole order and post-Newtonian terms up to monopole order \cite{thorne80}), one finds that the GW power $\EGW \equiv d E_{\text{GW}}/dt$ for an axisymmetric body reads\footnote{For $\gamma > 1$, the radiation energy $\EGW$ \eqref{eq:h0} can in fact be negative, entailing a scenario wherein negative energy is radiated away from the star \cite{sexl66}. This feature is often excluded on stability grounds \cite{ppn2}, so we consider $\gamma \leq 1$ here.}
\begin{widetext}
\begin{equation} \label{eq:h0}
\begin{aligned}
\dot{E}^{\text{ST}}_{\text{GW}} =& \frac {1536 \pi^6 G \nu^2} {c^5} \left\{ \frac {2\left(1 + \gamma \right)} {15} \nu^4 \epsilon^2 I_{0}^2  +  \pi^2 \left( 1 - \gamma \right) \left\{ \int dr d \theta \delta \rho(r,\theta) r^2 \sin\theta \left[ r^2 \nu^2 +  \left(6 \gamma -2 \right)  U(r) \right] \right\}^{2} \right\},
\end{aligned}
\end{equation}
\end{widetext}
where $I_{0}$ is the moment of inertia, and we have introduced the gravitational ellipticity $\epsilon$ \cite{mmra11},
\begin{equation} \label{eq:epsilon}
\epsilon = \pi I_{0}^{-1} \int_{V} dr d \theta \delta \rho(r,\theta) r^4 \sin\theta \left( 1 - 3 \cos^2\theta \right),
\end{equation}
as a convenient proxy for the tensor quadrupole moment contribution \cite{cut02,brady98,aasi14}. In expression \eqref{eq:h0}, the first term represents the tensor quadrupole contribution, which has the same form as the leading term in GR modulo a weighting by a $1 + \gamma$ term, which enters because the scalar field couples to the stress-energy tensor \cite{scalarten}. The remaining two terms are contributed by the scalar field directly and represent the Newtonian quadrupole and post-Newtonian monopole pieces, respectively. Clearly, for any given source candidate\footnote{From an experimental point of view, it is important to note that there does not exist an orientation for an interferometer relative to the source which is optimal for detection of both the tensor and scalar modes simultaneously. In particular, the optimal state for tensor mode detection is obtained when the wobble angle is $\pi/2$ and the angle made between the angular momentum and line of sight vectors, with respect to the observer, is zero \cite{jaran}. For this orientation, the detectability of scalar modes is effectively weakened by a factor $2$ \cite{orient1,orient2}. Oscillation eigenfrequencies are also dependent on the particulars of the scalar field \cite{sotani} (cf. Ref. \cite{suveig}), which may affect the detectability of the source.}, the major uncertainties in expression \eqref{eq:h0} are the magnitude of the density deformation $\delta \rho$ [which is implicitly a function of the PPN and magnetic field parameters through \eqref{eq:ppnrho}] and the Eddington parameter $\gamma$. Note that since the monopole piece emerges at the post-Newtonian level, $\EGW^{\text{mono}}$ scales with the square of the dimensionless compactness $\mathcal{C} = G M / c^2 R_{\star}$ of the star; cf. $\mathcal{C} \approx 0.2$ for a neutron star while $\mathcal{C} \lesssim 10^{-4}$ for a white dwarf \cite{tovx}. In general, the ratio of the monopole to quadrupole contributions to \eqref{eq:h0} can be estimated as
\begin{equation} \label{eq:ratio}
\frac{\EGW^{\text{mono}}}{\EGW^{\text{quad}}} \sim 10^{10} \left(1 - \gamma \right) \left( \frac {\mathcal{C}} {0.2} \right)^{2} \left( \frac {\nu} {100 \text{ Hz}} \right)^{-4} \left( \frac{ \rS} {10^{6} \text{ cm}} \right)^{-4},
\end{equation}
implying that monopolar radiation is likely to dominate, even for rapidly rotating stars $\nu \lesssim 1 \text{ kHz}$, unless $1-\gamma$ is sufficiently small. 

\begin{figure*}
\includegraphics[width=\textwidth]{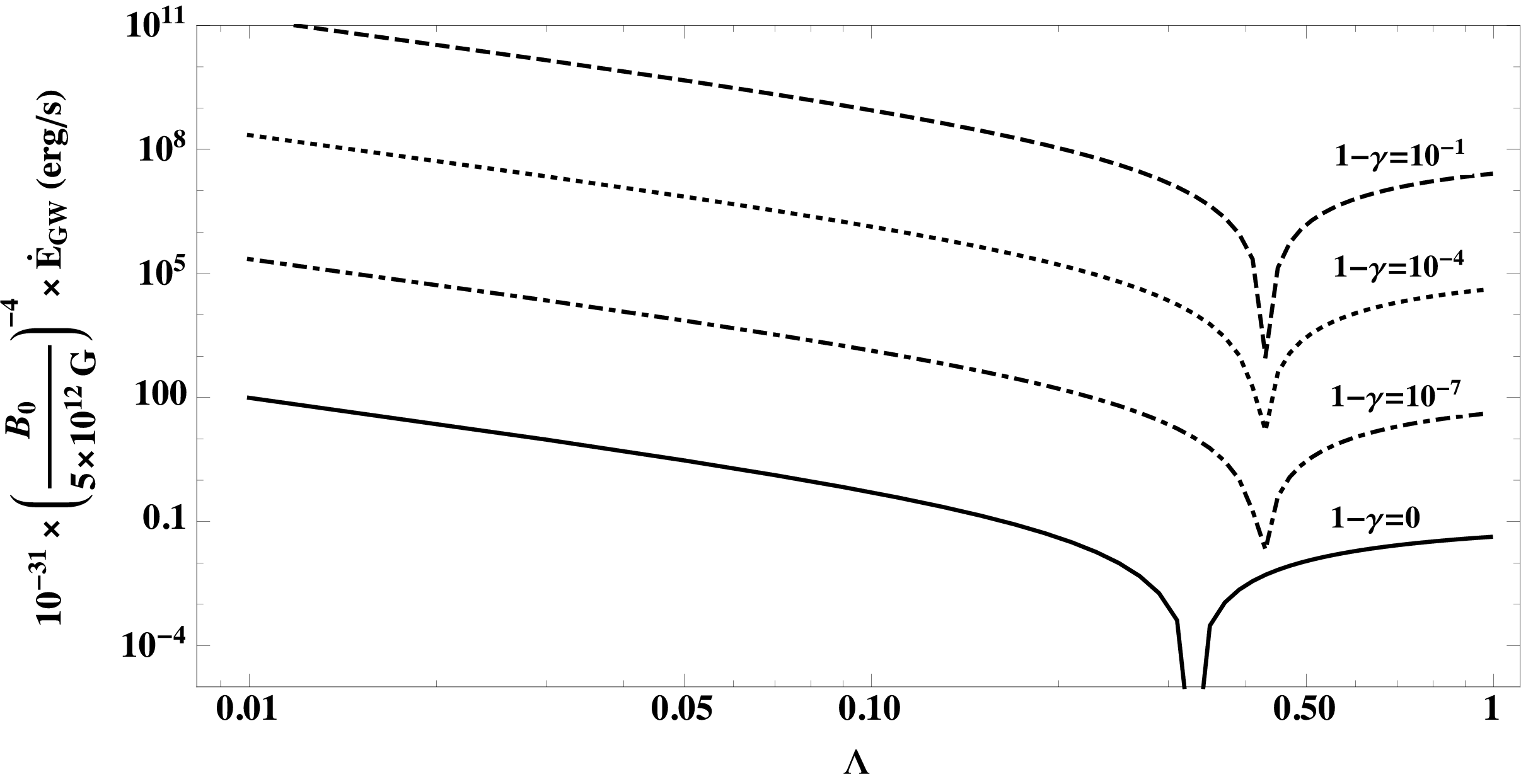}
\caption{Power radiated in GWs $\EGW$ as a function of $\Lambda$ for $1-\gamma =0$ (GR value; solid curve), $1-\gamma = 10^{-7}$ (dotdashed curve), $1-\gamma = 10^{-4}$ (dotted curve), and $1-\gamma = 10^{-1}$ (dashed curve), with $\beta = \beta_{2} = \beta_{4} = 1$ (GR values), spin frequency $\nu = 100 \text{ Hz}$, stellar radius $R_{\star} = 10^{6} \text{ cm}$ and mass $M_{\star} = 1.4 M_{\odot}$. \label{epsgamma}
}
\end{figure*}

In Figure \ref{epsgamma} we plot the GW power $\EGW^{\text{ST}}$ \eqref{eq:h0} as a function of poloidal-to-toroidal field strength $\Lambda$ for the dipolar magnetic field configuration detailed in the previous section for various values of $\gamma$ (different curves). Note that the star is prolate $(\epsilon < 0)$ for strongly toroidal $(\Lambda \rightarrow 0)$ fields and oblate $(\epsilon > 0)$ for strongly poloidal $(\Lambda \rightarrow 1)$ fields \cite{cut02,mmra11}. This prolate-oblate switch occurs because integrals over $\delta \rho$ have a local minimum around $\Lambda \approx 0.38$, the exact location varying slightly as a function of the PPN parameters (cf. Fig. \ref{rho10}). This sign-flip manifests as the sharp dips in the values of $\EGW$ that are seen around $\Lambda \approx 0.38$ in Fig. \ref{epsgamma} across all values of $\gamma$. We see that the monopole breathing mode dominates even when $1 - \gamma \sim 10^{-7}$ (dotdashed curve), as expected from expression \eqref{eq:ratio}. Strong deviations from the GR prediction (solid curve) are apparent; $\EGW^{\text{ST}}/\EGW^{\text{GR}} = 9.9 \times 10^{2}$ for $\Lambda = 1$, $1-\gamma = 10^{-7}$, and $\nu = 100 \text{ Hz}$, for example.  

\begin{figure}[h]
\includegraphics[width=0.473\textwidth]{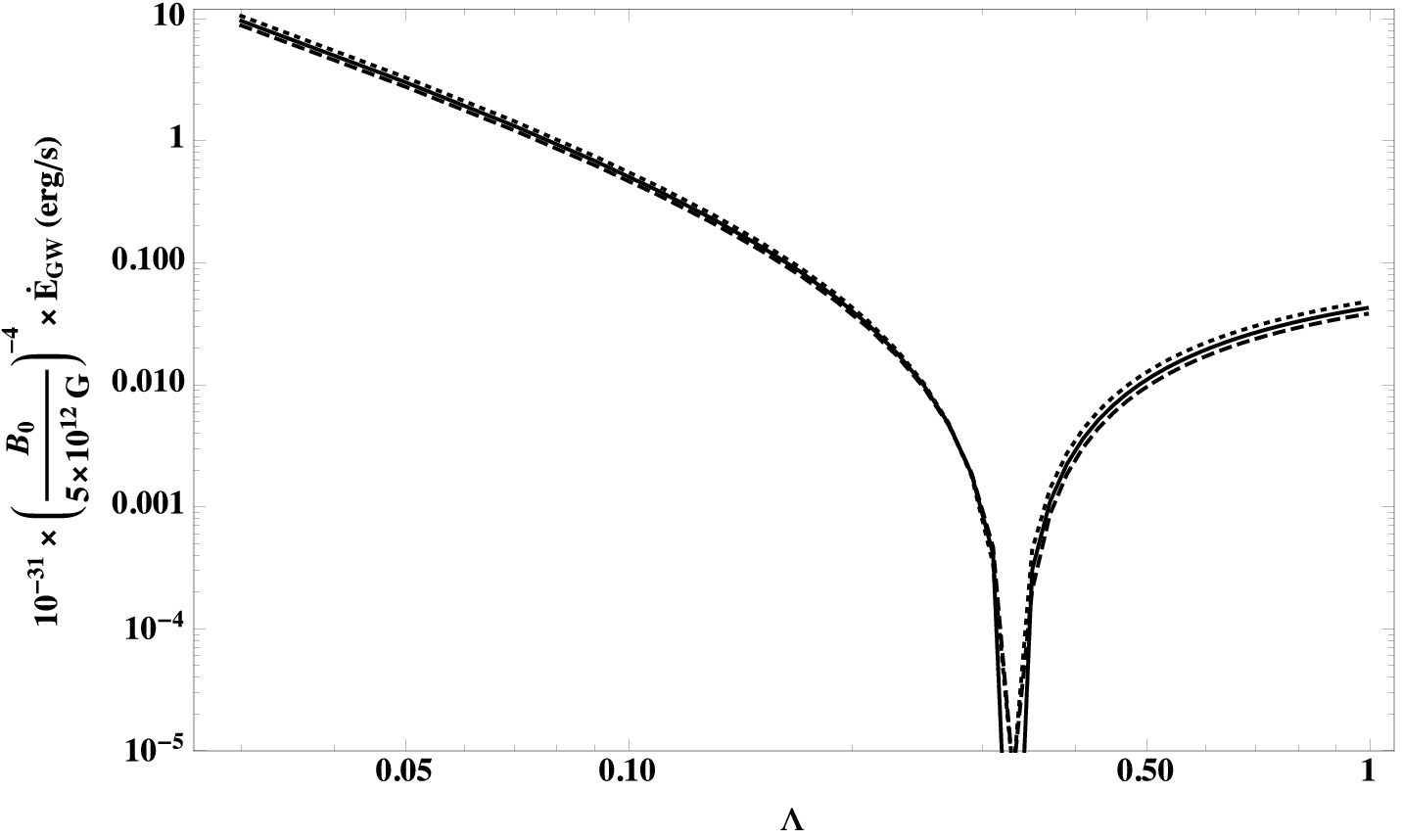}
\caption{Power radiated in GWs $\EGW$ as a function of $\Lambda$ for $\beta =1$ (GR value; solid curve), $\beta = 0.8$ (dashed curve), and $\beta = 1.2$ (dotted curve), with $\gamma = \beta_{2} = \beta_{4} = 1$ (GR values), spin frequency $\nu = 100 \text{ Hz}$, stellar radius $R_{\star} = 10^{6} \text{ cm}$ and mass $M_{\star} = 1.4 M_{\odot}$. \label{epsbeta}
}
\end{figure}

\begin{figure}[h]
\includegraphics[width=0.473\textwidth]{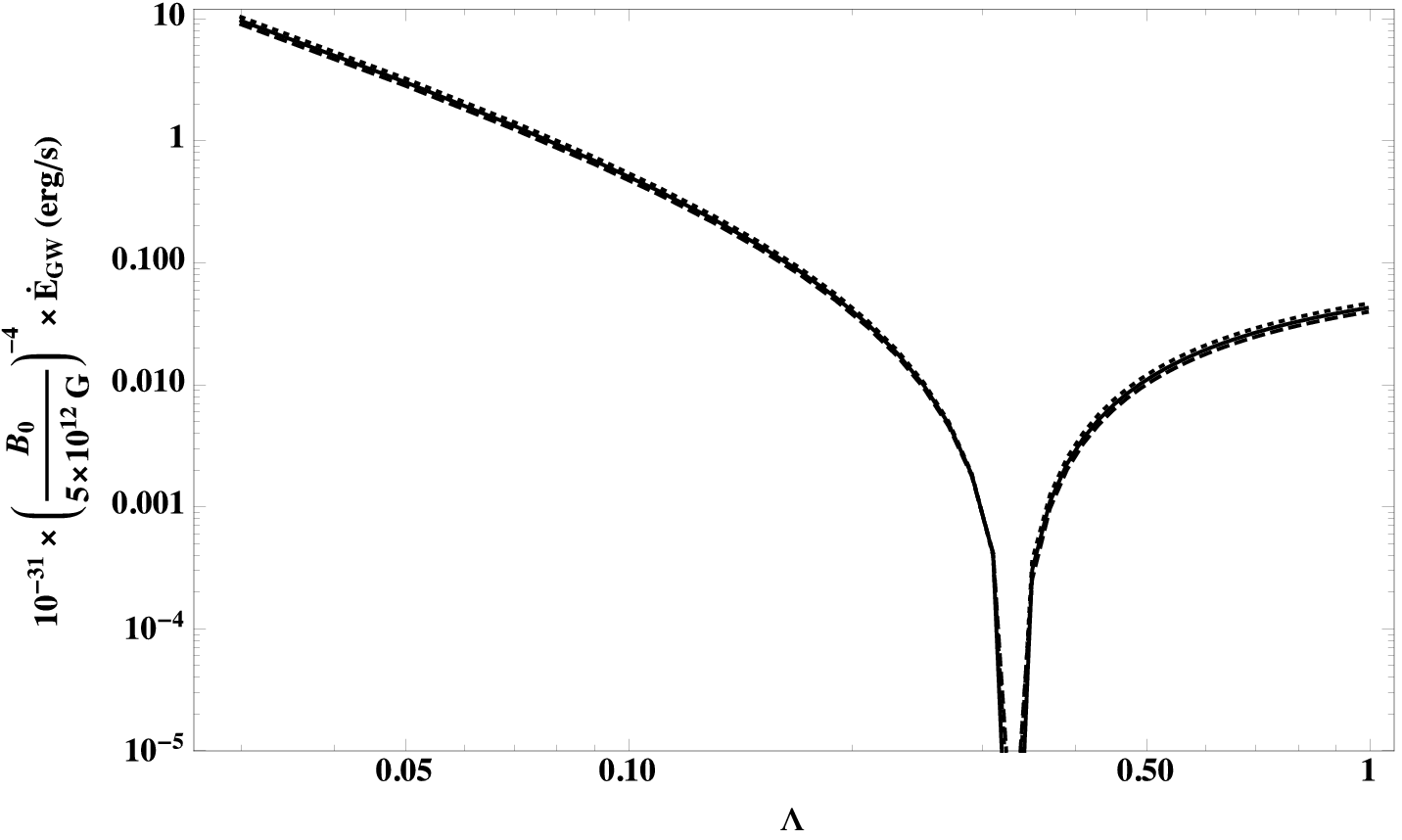}
\caption{Power radiated in GWs $\EGW$ as a function of $\Lambda$ for $\beta_{2} =1$ (GR value; solid curve), $\beta_{2} = 0.8$ (dashed curve), and $\beta_{2} = 1.2$ (dotted curve), with $\gamma = \beta = \beta_{4} = 1$ (GR values), spin frequency $\nu = 100 \text{ Hz}$, stellar radius $R_{\star} = 10^{6} \text{ cm}$ and mass $M_{\star} = 1.4 M_{\odot}$. \label{epsbeta2}
}
\end{figure}

\begin{figure}[h]
\includegraphics[width=0.473\textwidth]{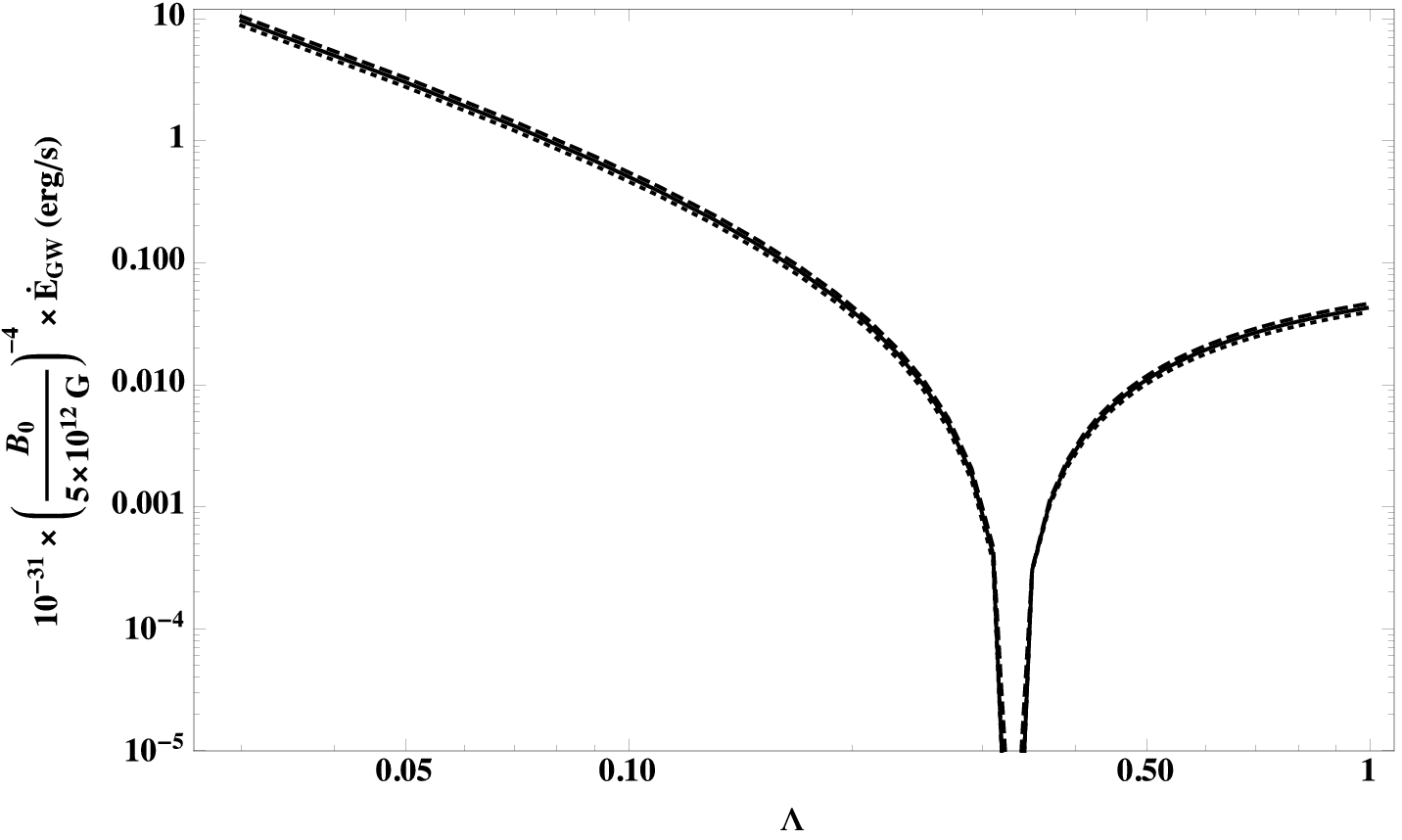}
\caption{Power radiated in GWs $\EGW$ as a function of $\Lambda$ for $\beta_{4} =1$ (GR value; solid curve), $\beta_{4} = 0.8$ (dashed curve), and $\beta_{4} = 1.2$ (dotted curve), with $\gamma = \beta = \beta_{2} = 1$ (GR values), spin frequency $\nu = 100 \text{ Hz}$, stellar radius $R_{\star} = 10^{6} \text{ cm}$ and mass $M_{\star} = 1.4 M_{\odot}$. \label{epsbeta4}
}
\end{figure}

Figures \ref{epsbeta}, \ref{epsbeta2}, and \ref{epsbeta4} are similar to Fig. \ref{epsgamma} except that the PPN parameters $\beta, \beta_{2}$, and $\beta_{4}$, respectively, are varied instead of $\gamma$, which is kept at its GR value. Since we set $\gamma=1$, there is no monopole contribution to the radiation energy $\EGW$. As such, Figs. \ref{epsbeta}--\ref{epsbeta4} effectively illustrate how the gravitational ellipticity \eqref{eq:epsilon} varies as a function of the non-$\gamma$ PPN parameters. We see that, even for $\mathcal{O}(10^{-1})$ departures in these parameters, the overall radiation energy is largely unchanged; $\EGW(\beta=1.2)/\EGW(\beta=0.8) = 1.26$ for $\Lambda = 1$ and $\EGW(\beta_{4}=0.8)/\EGW(\beta_{4}=1.2) = 1.19$ for $\Lambda = 10^{-2}$, for example. Small departures in $\EGW$ even for significant changes in $\beta, \beta_{2}$, or $\beta_{4}$ suggest that, aside from the possibility of monopolar radiation, leading-order modified gravity terms are unlikely to significantly affect the detectability of a magnetically deformed neutron star from a GW standpoint.


\subsection{Pulsar constraints}

As explored in the previous section, even fractional departures from unity in the value of $\gamma$ can dramatically adjust the GW luminosity because of monopolar radiation \eqref{eq:ratio}; see Fig. \ref{epsgamma}. This allows us to place $\Lambda$-dependent constraints on $\gamma$ in the strong field regime $R \sim R_{\text{NS}}$ from the observational upper limits of $\EGW$ from various neutron stars. {Recently, the LIGO collaboration} presented upper limits for $\EGW$ from the {first science runs of Advanced LIGO for a number of pulsars} \cite{aasi14}. In this study, {eleven} young pulsars were labeled as ``high value" because of their high spindown luminosities (which suggests significant radiation, through GWs or otherwise). We consider these {eleven} pulsars here.


Assuming the star acts as an orthogonal rotator, one can estimate the equatorial magnetic field strength $B_{\star}$ for a pulsar by bounding the electromagnetic braking energy by the rotational kinetic energy loss (e.g. \cite{tovx}), viz.
\begin{equation} \label{eq:charb}
B_{\star} \gtrsim \left| \frac {3 c^3 I_{0}} {8 \pi^2 R_{\star}^{6}}  \frac {\dot{\nu}} {\nu^3} \right|^{1/2}.
\end{equation}
By fixing the parameter $B_{0} = B_{\star}$ in our models to be the conservative lower limit of expression \eqref{eq:charb}, we can compare our calculations for $\EGW$ \eqref{eq:h0} with the observational upper limits presented {in Ref.} \cite{aasi14}. The results are shown in Table \ref{tab:tablims}. In particular, we place bounds on the parameter $\gamma$ by demanding that expression \eqref{eq:h0} takes a value less than the experimental limits, assuming canonical neutron star parameters $M_{\star} = 1.4 M_{\odot}$ and $R_{\star} = 10^{6} \text{ cm}$. We consider two field configurations, one which is purely poloidal  $(\Lambda = 1)$ and another which is predominantly toroidal $(\Lambda = 10^{-2})$.  

We see that for the most conservative models wherein toroidal field contributions are negligible, relatively weak constraints are placed on the parameter $\gamma$; for the Vela {and Crab} pulsars we obtain the bounds $1 - \gamma \leq 4.2 \times 10^{-3}$ {and $1- \gamma \leq 8.4 \times 10^{-3}$, respectively}. We conclude therefore that, from magnetic deformations by purely poloidal fields with characteristic strength $B_{\star}$ given by the lower bound of \eqref{eq:charb}, constraints on modified gravity parameters are modest. Note, however, that taking characteristic field strengths $B_{\star}$ larger than the minimum permitted by \eqref{eq:charb} will place stronger constraints as $\EGW \propto \delta\rho^2 \propto B_{\star}^4$. A more compact stellar model with greater $\mathcal{C}$ or smaller radius would also yield stronger constraints \eqref{eq:ratio}. Moreover, models with a strong toroidal field $(\Lambda = 10^{-2})$ allow us to place constraints on $\gamma$ that are competitive with Solar system experiments \eqref{eq:solar1}, even for the lower limit \eqref{eq:charb}. For Vela, we find that $1-\gamma \leq 8.0 \times 10^{-7}$ is necessary in order for the theory to be consistent with observation, which is stronger than \eqref{eq:solar1} by over an order of magnitude and, importantly, applies to the strong field regime $R \sim R_{\text{NS}}$; cf Fig. \ref{gammabeta}.

We close this section by recalling that, as mentioned in the introduction, electromagnetic observations, such as measurements of resonant cyclotron \cite{cyc1} or X-ray pulse fraction \cite{xray} features, suggest that some neutron stars have local magnetic field strengths much greater than the `global' strengths inferred from their spin down limits \eqref{eq:charb}; $\gtrsim$ 3 orders of magnitude discrepancies are found for 1E $1207.4$-$5209$ \cite{psrb}, for example. It has been shown that this contrariety can be naturally explained through strong internal toroidal fields \cite{gep03,gepvig14}.  Stability and evolutionary studies of neutron star magnetic fields \cite{stelev1,stelev2}, simulations of core-collapse supernova \cite{core1}, and models of surface temperature anisotropies \cite{dany1} further support the suggestion that some neutron stars, especially young ones, admit strong toroidal fields. As such, values $\Lambda \lesssim 10^{-2}$ or $B_{0} \gg B_{\star}$ from \eqref{eq:charb} are perhaps not unrealistic in reality. 


\begin{table*}
\caption{Comparison of our models with observational limits of {eleven} selected pulsars. The characteristic magnetic field strength $B_{0}$ is set by spindown, $B_{0} = B_{\star}$, where $B_{\star}$ is estimated using \eqref{eq:charb} and the spin frequency data $(\nu,\dot{\nu})$ given in Ref. \cite{aasi14}. The fourth column shows the observational upper limits from a Bayesian analysis on $\EGW$ from Ref. \cite{aasi14} at the $95\%$ confidence level. The fifth and sixth columns shows bounds on $1-\gamma$ obtained by requiring that $\EGW^{\text{ST}} \leq \EGW^{95\%}$, with $\EGW^{\text{ST}}$ calculated from \eqref{eq:h0} using \eqref{eq:ppnrho} for stars with purely poloidal $(\Lambda = 1)$ and primarily toroidal $(\Lambda = 10^{-2})$ magnetic fields, respectively. We have assumed that all PPN parameters except for $\gamma$ take their GR values in calculating these estimates; cf. Figs \ref{epsbeta}--\ref{epsbeta4}. Note that a $-$ indicates that even an $\mathcal{O}(1)$ value for $1-\gamma$ does not allow $\EGW^{\text{ST}} > \EGW^{95\%}$. We set $R_{\star} = 10^{6} \text{ cm}$ and $M_{\star} = 1.4 M_{\odot}$.}
  \begin{tabular}{llcccc}
  \hline
Pulsar & $B_{\star}$ & $\nu$ &  Observational limit $\EGW^{95\%}$ & $1-\gamma$& $1-\gamma$ \\
 &$(10^{12} \text{ G})$ & (Hz)  & ($10^{35}$ $\text{erg} / \text{s}$) & $(\Lambda = 1)$ & $(\Lambda = 10^{-2})$  \\
\hline
J$0205$+$6449$ & $3.60$ & $15.2$ & $17.2$ & $\leq 1.2 \times 10^{-1}$ & $\leq 1.3 \times 10^{-5}$ \\
J$0534$+$2200$ (Crab) & $3.75$ & $29.7$  & $9.61$ & $\leq 8.4 \times 10^{-3}$ &  $\leq 1.5 \times 10^{-6}$   \\
J$0835$-$4510$ (Vela) & $3.62$  & $11.2$ & $0.617$ & $\leq 4.2 \times 10^{-3} $  & $\leq 8.0 \times 10^{-7}$   \\
J$1302$-$6350$ & $0.337$ & $20.9$ & $8.30$ & $-$ & $\leq 5.2 \times 10^{-2}$ \\
J$1809$-$1917$ & $1.46$ & $12.1$ & $110$ & $-$ & $\leq 4.7 \times 10^{-3}$ \\
J$1813$-$1246$ & $0.956$  & $20.8$ & $3.15$ & $-$  & $\leq 2.6 \times 10^{-4}$   \\
J$1826$-$1256$ & $3.74$ & $9.05$ & $161$ & $-$ & $\leq 3.4 \times 10^{-4}$ \\
J$1928$+$1746$ & $0.974$ & $14.6$ & $147$ & $-$ & $\leq 2.4 \times 10^{-2}$ \\
J$1952$+$3252$ (CTB 80)      & $0.472$  & $25.3$ & $7.66$ & $-$ & $\leq 2.1 \times 10^{-2}$   \\
J$2043$+$2740$ & $3.87$ & $10.4$ & $23.1$ & $-$ & $\leq 2.7 \times 10^{-5}$ \\
J$2229$+$6114$ & $2.03$ & $19.4$ & $5.12$ & $-$ & $\leq 2.3 \times 10^{-5}$ \\

\hline
\end{tabular}
\label{tab:tablims}
\end{table*}

\section{Discussion}

In this paper we develop a formalism to study the properties of magnetised, post-Newtonian stars in metric theories of gravity which introduce modifications to GR in the strong field regime. The formalism has the benefit that it can, in principle, incorporate arbitrary magnetic field and fluid configurations \cite{mmra11,lasky13,mast16} and many (see Footnote 1) metric theories of gravity \cite{ppnX,ppnNI,ppn1,ppn2} simultaneously. We further estimate the power in GWs continuously radiated by a magnetically deformed neutron star  in terms of both the PPN and magnetic field parameters; see Figs. \ref{epsgamma}--\ref{epsbeta4}. For theories of gravity which permit it, we find that the monopole contribution to the continuous GW luminosity $\EGW$ \eqref{eq:h0} can easily dominate over the usual quadrupole contributions unless the Eddington parameter $\gamma$ is small; see expression \eqref{eq:ratio}. A comparison with observational upper limits on $\EGW$ for {eleven} pulsars from LIGO and Virgo data is used to place constraints on the Eddington parameter $\gamma$; see Tab. \ref{tab:tablims}.

For purely poloidal magnetic fields ($\Lambda = 1$ in our terminology), the obtained constraints are relatively weak; we find $1-\gamma \leq 4.2 \times 10^{-3}$ is necessary in order for theoretical predictions to be consistent with observations of the Vela pulsar \cite{aasi14}. However, supposing the star to possess a strong toroidal field, as predicted from a variety of stability and evolutionary scenarios \cite{stelev1,stelev2,core1,dany1}, the derived constraints are at least an order of magnitude stronger than Solar system constraints \eqref{eq:solar1} \cite{cassini,lunar}. In particular, for models where $99\%$ of the magnetic energy is contained within a toroidal field ($\Lambda = 10^{-2}$), we find that limits on the GW luminosity of Vela require $1-\gamma \leq 8.0 \times 10^{-7}$. Furthermore, although the energy radiation formula \eqref{eq:h0} that we use is particular to scalar-tensor theories of gravity, the constraints presented in Table \ref{tab:tablims} are relatively generic as they involve the Eddington parameter $\gamma$, which is not scalar-tensor specific. Indeed, any theory which predicts monopolar radiation will necessarily have a coefficient scaling the monopole term which is proportional to some function of $1-\gamma$ (and possibly other PPN parameters) which tends to zero in the GR limit \cite{sexl66}. Aside from monopolar radiation, however, we find little departure from the GR value of $\EGW$ even when the other non-$\gamma$ PPN parameters deviate significantly from unity; see Figs. \ref{epsbeta}--\ref{epsbeta4}. This implies that, in the absence of monopolar radiation, a detection of continuous GWs from neutron stars would more likely teach us about the properties of dense matter and not about strong gravity because leading-order GR corrections have negligible consequences as far as the GW power is concerned (cf. Ref. \cite{sotani}).


Even in the post-Newtonian theory, we find that density perturbations are proportional to the magnetic energy, i.e. $\delta \rho \propto B_{0}^2$. This implies that the relative deformation $\delta \rho / \rho$ might be much larger in a magnetar than for the pulsars discussed in Sec. V. B \cite{duncan,laskrev,mmra11}. However, expression \eqref{eq:charb} implies a strong magnetic field exists precisely when the star has a low spin frequency, i.e. $B_{\star} \propto \nu^{-3}$, which generally implies, at least in GR, that the GW luminosity of an older magnetar is relatively low. However, since monopolar radiation is largely independent from the spin frequency of the star, one would expect that modified theories of gravity which permit breathing modes also predict very large GW luminosities for magnetars, especially ones with strong toroidal fields $\Lambda \lesssim 10^{-1}$. Future searches for continuous GWs from magnetars may therefore allow one to place constraints on $1-\gamma$ which are considerably stronger than those presented in Tab. \ref{tab:tablims}.


\section*{Acknowledgements}
{We thank the anonymous referee for their helpful suggestions.}




\end{document}